\documentclass[
 amsmath,amssymb,
]{revtex4-2}

\usepackage{natbib}
\usepackage{bm}
\usepackage{amsfonts}
\usepackage{amssymb}
\usepackage{amsmath}
\usepackage{amsthm}
\usepackage{mathrsfs}
\usepackage{graphicx}
\usepackage{physics}
\usepackage{bbm}

\begin{document}

\title{Wave equation of massless particles of arbitrary helicity}

\author{Abraham Lozada}
 \email{abelozada@gmail.com}
\affiliation{ 
Escuela de F\'isica, Facultad de Ciencias, Universidad Central de Venezuela, Caracas, Venezuela
}

\author{S. Tabban}
 \email{sm.tabban@uniandes.edu.co}
\affiliation{ 
Departamento de F\'isica, Universidad de los Andes, Bogot\'a,
Colombia
}

\date{\today}

\begin{abstract}
In this work, we derive from first principles the relativistic wave equation of massless particles of arbitrary helicity. We start from unitary projective irreducible representations of the restricted Poincar\'e group.  We define a weaker notion of localization and find, in particular, a position operator for any massless particle of arbitrary helicity. Therefore, having the position representation for these particles, we obtain the wave equations in this representation. We compare our results with previous findings in the literature.
\end{abstract}


\maketitle

\section{Introduction}

Since the beginning of quantum mechanics, the definition of the position observable of a particle has been an important issue. For instance, without it, we cannot
define Heisenberg uncertainty relations and constructing other observables that are functions of the position. For the case of elementary
massive relativistic particles,  there exists a notion of localization \cite{Newton1949,Wightman1962}. However, in the case of massless particles of  non-zero helicity, the above notion of localization does not exist \cite{Newton1949,Wightman1962}. In particular, a very important case, the photon is not localizable (the only massless particle experimentally found).

 This property of massless particles, with  non-zero helicity,
avoids the existence of position representations for them. Although quantum mechanics does not depend on the representation, the practical
importance of the position representation is unquestionable. In fact, the amount of old and new publications (see, e.g,   \cite{Birula1996,Birula2019,Birula2020}) support this statement. In particular, see some recent publications in this journal \cite{Babaei2017,Kiessling2018}. The wave function of the photon, in the position representation, is important because it is a direct and simpler theoretical approach to experiments involving a constant number of photons. For example, the ones with one photon, and others involving entanglement and quantum information with a fixed number of photons.

It is noteworthy, that in the literature (see, e.g,   \cite{Birula1996,Birula2019,Birula2020,Babaei2017,Kiessling2018}), it is considered a position representation for massless particles in spite of the non localization results. Some of these works do not elaborate about this issue. On the other hand, they postulate wave equations, for instance, in the case of the photon, analogous to Maxwell's equations. Nevertheless, the wave equations should be obtained from the representations of massless particles with arbitrary helicity. That is, they should not be postulated. Moreover, postulating the wave equations could lead  to fundamental problems like superposition of particles with different helicity.
For instance,  it is impossible to prepare a one-particle state representing a photon of helicity $+1$ with some probability \textrm{p} ($0< \mbox{\textrm{p}} < 1$) and a photon of helicity $-1$ with probability $1-$\textrm{p}. That is, helicity is a superselection rule. However, sometimes, it is not taken into account explicitly because the photon is thought as being its own antiparticle (see, e.g.,   \cite{Weinberg1995}).

The aim of this paper is to obtain the wave equation (in the position representation) of a massless particle with arbitrary fixed helicity. Therefore, we have to define in what sense there exists the position representation. In order to do that, we consider a trivial  notion of localization weaker than the usual one \cite{Newton1949,Wightman1962}. Having the notion of localization, we start from the projective unitary irreducible representation of the Poncar\'e group that defines the massless particle. Then, from this representation, we \textit{derive} the wave equation of the elementary particle with definite helicity. In contraposition with previous results, we distinguish between particles with helicity $+\lambda$ and $-\lambda$ (because they are different elementary particles, like the electron and the positron).

In this paper, we will use Planck units ($c = 1 = \hbar$). This work is organized as follows. In section II, we define the notion of localization as a restriction of the Newton-Wigner-Wightman (NWW) axioms. In section III, we remind the irreducibles representations of massless particles, of arbitrary helicity, that we are going to consider. In section IV, we prove that massless particles are localizable, in the sense of section II, and we derive, from the representations of section III, the wave equations in the position representation. In section V, we consider the possibility of describing the irreducible representations, used in section III and IV, as restrictions of reducible representations. Finally, in section VI, we make some remarks and conclusions.

\section{Localization}

Our problem is to obtain the form of the wave equation of massless particles in the position representation. Therefore, in this section, we define what we understand by a position representation.

We consider the NWW  axioms \cite{Newton1949,Wightman1962} to define a position representation. However, we do not use all the axioms, because of the non localization results for massless particles. In fact, we eliminate the part from axiom V, of Wightman's paper \cite{Wightman1962}, that has to do with the action of the rotations group in the position representation. That is, we only
look for an imprimitivity system for the translation subgroup instead of an imprimitivity system for the euclidean group. Then, the NWW localization axioms are restricted now:
\begin{itemize}
 \item[I.] For every Borel set $S$ of $\mathbb{R}^3$, there is a projection operator $E(S)$ whose expectation value is the probability of finding
the system in $S$.
 \item[II.] $E(S_1 \cap S_2) = E(S_1) E(S_2)$.
 \item[III.]$E(S_1 \cup S_2) = E(S_1)+ E(S_2) - E(S_1 \cap S_2)$. If $S_i$, $ \; i \in 1, 2 , \cdots$ are disjoint Borel then \\ $E(S_1 \cup S_2 \cup \cdots) = \sum_{i = 1}^\infty E(S_i).$
 \item[IV.] $E(\mathbb{R} ^3) = \mathbb{I}$.
 \item[V.] $E(\mathbbm{1} S + \vec{a}) =  U(\vec{a},\mathbbm{1}) E(S) U^{-1}(\vec{a},\mathbbm{1})$, where  $\vec{a} \in \mathbb{R}^3$, with $\mathbb{R}^3$ considered as the position space which is an affine space too, and $ U(\vec{a},\mathbbm{1})$ is a unitary representation of the 3-dimensional translation group belonging to the physical system in question.
 
\end{itemize}

The first four axioms are the same as Wightman \cite{Wightman1962}. In the axiom V, the symbol $(0,\mathbbm{1})$ is the identity of the Euclidean group, therefore, $ U(\vec{a},\mathbbm{1})$ 
is the restriction of the unitary representation of the universal covering of the restrictive Poincar\'e group, $\widetilde{P}_+^\uparrow$, that defines the physical system in question, to the spatial translation group. 

Our trivial axioms are satisfied for massive particles since they are localized in the stronger NWW sense. Also, because we are going to prove, in section \ref{section:wave_equation}, that massless particles are localizable in our sense, we conclude that all the elementary particles are localized in the sense of our axioms.

Let us note that axiom V is a reformulation, see Mackey \cite{Mackey1968}, of the Weyl canonical commutation relations. Therefore, we are only asking for satisfying  Weyl canonical commutation relations. It is trivial, in comparison with NWW axioms, but it is preferable to pay this price that no having a position representation.

\section{Massless particles}

In this section, we define the kind of massless particle that we are going to consider.

According to Wigner theorem \cite{Bogolubov1975,Weinberg1995}, an elementary particle corresponds to a projective unitary irreducible representation of the
restrictive Poincar\'e group\cite{Bogolubov1975,Weinberg1995}. These projective representations are generated by true representations of $\widetilde{P}_+^\uparrow$\cite{Bogolubov1975,Weinberg1995}. In the case of massless particles,  discrete helicity, positive energy, these representations act in the Hilbert space of momentum $L^2
\left(X_0,\frac{d^3p}{|\vec{p}|}\right)$, where $\frac{d^3p}{|\vec{p}|}$ is the Lorentz invariant measure and $$ X_0 :\left\lbrace p \in
\mathbb{R} ^{3+1} \; | \; p^\mu  p_\mu=  0, \; p^0 > 0  \right\rbrace$$ is the forward light cone in momentum space. They are given by \cite{Lledo2003,Flato1983}:
\begin{equation}\label{eq_massless_irrep}
\qty[U(a, A)]\phi(p) =  e^{-i p^\mu a_\mu} e^{i\lambda  \theta(A,p)}  \phi \qty( \Lambda^{-1}(A) p),
\end{equation}
where $\phi \in L^2\left(X_0,\frac{d^3p}{|\vec{p}|}\right)$, $a \in \mathbb{R} ^{3+1}$, $p^\mu a_\mu =  p^0a^0 - p^1a^1 - p^2a^2 - p^3a^3$, $\lambda$ is the integer or half-integer helicity, $\theta(A,p) \in \mathbb{R}$ is a scalar that, besides $p$, depends\cite{Flato1983} on $A \in SL(2,\mathbb{C})$  and  $\Lambda$ is the unique Lorentz transformation associated to $A$.  

In the case of the photon, $\lambda = 1$ or $\lambda = -1$. However, because the representations with  $\lambda$ and $-\lambda$ are inequivalent, we have to distinguish between the elementary particles associated to each representation. In particular, we have two kinds of photons (photons with helicity $+1$ and photons with helicity $-1$). This has as a consequence that the evolution equation of a particle with defined helicity is not invariant under parity. In the case of theories that are invariants under parity, this is not a problem. In fact, what is invariant under parity is the quantum field theory (QFT) from which the particles are associated. 

In view of what we have said, in consistency with the fact that a particle with defined helicity is not invariant under parity,  we cannot think of $\phi_\pm$ as states of photons with helicity $\pm 1$, respectively, and writing a general one-photon state as
\begin{equation}
  \phi =  \alpha_+ \phi_+ + \alpha_- \phi_-,
\end{equation}
where $|\alpha_+|^2 + |\alpha_-|^2 = 1$, with $|\alpha_+| > 0$ and $|\alpha_-| > 0$.

\section{Wave equation}\label{section:wave_equation}

In this section, we are going to prove that all the massless particles with any helicity, are localizable in the sense of the axioms I-V. Additionally, as a consequence, we derive the wave equation of massless particles from the representation given by \eqref{eq_massless_irrep}. It is important to emphasize that we do not postulate wave equations, our equations are a direct consequence of the axioms I-V and the representations given by \eqref{eq_massless_irrep}.

In order to prove that all the massless particles are localizable, we remind that the Schrödinger representation (position and momentum operators) is equivalent to the Weyl canonical commutation relations, also equivalent to an imprimitivity system of the translation group, as we pointed out before. Then, it is enough to show the existence of an unitary equivalent representation to \eqref{eq_massless_irrep} that admits the usual Schrödinger representation of position and momentum. This is what we do as follows.

First, to find the position representation, let us consider the unitary transformation
\begin{equation}\label{eq_V_transformation}
V: L^2 \left(X_0,\frac{d^3p}{|\vec{p}|}\right) \rightarrow  L^2_p(\mathbb{R}^3),
\end{equation}
defined by
\begin{equation}
V \phi := \frac{1}{\sqrt{p^0}} \phi \equiv \widetilde{\psi}.
\end{equation}
Then, we apply the Fourier inverse transform defined, as usual, by
\begin{equation}\label{eq_FT}
 \psi(\vec{x})  =  \frac{1}{(2\pi)^{3/2}} \int \widetilde{\psi}(\vec{p}) e^{ i \vec{x}.\vec{p}} d^3p,
\end{equation}
which is well-defined for all $\widetilde{\psi}(\vec{p}) \in L_p^1(\mathbb{R}^3) \cap L_p^2(\mathbb{R}^3)$. As it is well known, this transformation induces a unitary operator (Plancherel theorem):
\begin{equation}\label{eq_Fourier_transformation}
U_F^{-1}: L^2_p(\mathbb{R}^3) \rightarrow L^2_x(\mathbb{R}^3).
\end{equation}
From now on, we will use \eqref{eq_FT} as meaning the unitary operator. In $L^2_x(\mathbb{R}^3)$, the position observable and momentum observable are given, as usual, by the Schrödinger representation of the Weyl canonical
commutation relation. The projection operators associated to the position operator, the multiplicative by the characteristic functions associated to every Borel set, satisfy trivially the localization axioms I-V. So, the position representation exists for any massless particle. Any other representation its physically equivalent to this one (in the sense of the localization axioms I-V). In consequence, we understand what the position representation is for massless particles.

Now, we can consider the wave equation in the position representation $ L^2_x(\mathbb{R}^3)$. Starting from the projective irreducible representation of the  Poincar\'e group \eqref{eq_massless_irrep}, taking $A = \mathbbm{1}$ y $a= \dbinom{t}{\vec{0}} \equiv a_t \in \mathbb{R} ^{3+1}$, we obtain:
\begin{equation}
 \qty[U(a_t,\mathbbm{1})\phi](p)=e ^{-i p^0  t }\phi(p) \; ,
\end{equation}
where, we have used\cite{Lledo2003,Flato1983} $e^{i \lambda \theta(\mathbbm{1},p)} = 1$. This equation describes the evolution law of massless particles and gives the generator of temporal translation
\begin{equation}\label{eq_hamiltonian_evolution}
 \hat{H}\phi(p) =  \sqrt{\vec{p}^{\, 2}}\phi(p).
\end{equation}
And therefore, we obtain, as a consequence of the projective irreducible representation of the Poincar\'e group,  the relativistic evolution equation in momentum representation, for massless particles of any helicity:
\begin{equation}\label{eq_momentum_evolution}
 i \frac{\partial \phi(p,t)}{\partial t} =  \sqrt{\vec{p}^{\, 2}}\phi(p,t),
\end{equation}
where, $\phi \in \mathcal{D}(\hat{H}) \subset L^2\left(X_0,\frac{d^3p}{|\vec{p}|}\right)$, with $\mathcal{D}(\hat{H})$  the obvious domain of $\hat{H}$,  a dense subspace of $L^2\left(X_0,\frac{d^3p}{|\vec{p}|}\right)$ (as a function of the momentum observable, $\hat{H}$ is a self-adjoint operator).

The equation (\ref{eq_momentum_evolution}) has been obtained directly from \eqref{eq_massless_irrep}. Up to unitary equivalence, there is not other equation. This equation is not a postulate but a consequence of relation (\ref{eq_massless_irrep}). This is a completely relativistic equation, in the quantum sense, because relation (\ref{eq_massless_irrep}) defines relativistically, in the quantum sense, a massless particle of helicity $\lambda$ . 

In order to calculate the wave equation in the position representation, already defined, we transform the Hamiltonian operator $H=U_F^{-1}\hat{H}U_F$, that is:
\begin{equation}
\begin{array}{rcl}
(H\psi)(\vec{x}) & = &   \displaystyle {  \frac{1}{(2\pi)^{3/2}} \int \sqrt{(\vec{q})^2} \, \widetilde{\psi}(\vec{q}) e^{ i
\vec{x}.\vec{q}} d^3q } \\
& = & \displaystyle { \frac{1}{(2\pi)^3} \int \int \sqrt{(\vec{q})^2 } \, \psi(\vec{y}) e^{  i (\vec{x}-\vec{y}).\vec{q}}  d^3y  d^3q }.
\end{array}
\end{equation}
Hence, the wave equation, for any helicity $\lambda$, is given by:
\begin{equation}
\begin{array}{rcl}
 i   \dfrac{\partial \psi(\vec{x},t)}{\partial t} & = & (H \psi)(\vec{x},t)  \\ & = & \dfrac{1}{(2\pi)^3} \displaystyle { \int \int |\vec{q}| e^{
i(\vec{x}-\vec{y}).\vec{q}}\psi(\vec{y},t) d^3y d^3q },
 \end{array}
\end{equation}
which is equivalent to:
\begin{equation}\label{eq_position_evolution}
 i   \frac{\partial \psi(\vec{x},t)}{\partial t} = \sqrt{-\Delta}\psi(\vec{x},t),
\end{equation}
with $\Delta = \partial^2_{x_1} + \partial^2_{x_2} + \partial^2_{x_3}$, where, by definition:
\begin{equation}
 \sqrt{-\Delta}\psi(\vec{x}) = \frac{1}{(2\pi)^3} \int \int \sqrt{(\vec{q})^2}  e^{ i
(\vec{x}-\vec{y}).\vec{q}}\psi(\vec{y}) d^3y d^3q.
\end{equation}
Thus, we have obtained a pseudo-differential equation (in an adequate domain). We notice that the Hamiltonian is the unique square-root of the minus  self-adjoint  Laplacian, in $L^2_x(\mathbb{R}^3)$. Of course, this square-root is a well-defined self-adjoint operator (the energy observable in the position representation).

In this representation, the position operator is the multiplicative, therefore, the probability of finding the massless particle, given the normalized state $\psi(\vec{x},t)$, in a Borel $S$, in a time $t$, is 
\begin{equation}
    \displaystyle \int_S |\psi(\vec{x},t)|^2 d^3 x.
\end{equation}

Given the position operator in the position representation, we can calculate this operator in the momentum representation $L^2\left(X_0,\frac{d^3p}{|\vec{p}|} \right)$. As we know,  see \eqref{eq_V_transformation} and \eqref{eq_Fourier_transformation}, the position and momentum representations are connected by a unitary operator. Then, we obtain 
\begin{equation}\label{eq_position_operator}
    \left(\hat{\vec{X}} \phi \right)(p) = i \nabla_{\vec{p}} \phi(p) - i \dfrac{\vec{p}}{2|\vec{p}|^2}\phi(p).
\end{equation}
With the position operator given by \eqref{eq_position_operator}, we can calculate, as usual, the velocity operator. In order to do that, we go to the Heisenberg picture:
\begin{equation}
    \hat{\vec{X}}(t) = e^{iHt} \hat{\vec{X}} e^{-iHt}.
\end{equation}
The velocity operator is defined as the time derivative:
\begin{equation}
 \hat{\vec{V}}(t) = \dfrac{d}{dt} \hat{\vec{X}}(t) =  i [H,\hat{\vec{X}}(t)],
\end{equation}
that is,
\begin{equation}
 \hat{\vec{V}}(t) = \dfrac{\hat{\vec{p}}}{|\vec{p}|}.
\end{equation}
Therefore, the observable  $|\hat{\vec{V}}(t)| = \mathbbm{1}$,  as it should be, since every massless particle travels at the light velocity ($c=1$). Moreover, we have that $\dfrac{d}{dt} \hat{\vec{V}}(t) = 0$, as it should be too. All these properties are consequences of our position operator.

We notice that, despite all these nice properties of our position representation, we have the same problems of superluminal propagation \cite{Hegerfeldt1985,Thaller1992} as in the case of massive particles. This means that, after a Lorentz boost is applied to the wave function, this function has a non-local behavior in our case (as much as in the massive particle case). Also, as we have said, we have the non-local behavior in the rotation,  for massless particles with non-zero helicity, but this result is more preferable than the non localizability.

\section{Reducible representations and the Weyl equation}

In this section, we are going to study some projective reducible representations of the Poincar\'e group. We do that because, in the literature, reducible representations  are used for describing the photon and we want to compare with using only irreducible representations. Also, we want to show that considering irreducible representations, which is our point of view, as restriction of reducible representations could be technically advantageous.

Usually, see, e.g.,   \cite{Birula1996,Kiessling2018,Babaei2017}, a complexification of Maxwell type equations or the massless Dirac equation are proposed as the wave equation of the photon. This point of view has two drawbacks, one is the quantum relativistic invariance and the other is the possibility of mixing different physically nonequivalent representations. The first one arises because instead of considering a unitary representation of $\widetilde{\mathcal{P}}^\uparrow_+$, and deriving the wave equation, a wave equation (Maxwell or Dirac) is postulated. Therefore, to prove the relativistic invariance, a representation that is consistent with the proposed equation have to be built. However, to solve the issue of relativistic invariance, usually,  a Lie algebra representation of the Poincar\'e group is constructed. But, it is not necessarily enough (in fact,  the question of integrability of the algebra representation to a representation of the group must be elucidated). 
The second one appears because, if there exists a relativistic invariance, it is through a reducible representation.  Being a reducible representation, if we are not careful, we have a problem. In fact, we can construct, for instance, superposition of states of positive energy and negative energy, which not represent an elementary particle (as it happens with the Dirac equation for the electron).

Nevertheless, using reducible representations could be convenient, as we are going to show in this section, because we can remain in an irreducible representation,  as a restriction of the reducible one (like in the Dirac equation for the electron, eliminating the states of negative energy). To illustrate this point, we are going to show it, explicitly, for the case of a Weyl type wave equation. Also, this can be proven with the Maxwell type equation and, obviously, the massless Dirac equation (which is equivalent to two Weyl equations, each one with opposite helicity).

To begin with, let us construct the reducible representation, where the Weyl equation is acting, as a direct sum of two irreducible representations. Then, instead of given a representation where the Weyl equation is a consequence of time translation invariance, we postulate the Weyl equation (as it is done, see, e.g.,   \cite{Birula1996}, with the Maxwell type equations) and obtain the representation consistent with that equation. We shall see that any helicity is compatible with the Weyl equation.

Let $\qty(\mathcal{H}_1,\rho_1)$ and $ \qty(\mathcal{H}_2,\rho_2)$ be the irreducible representations of a massless particle, discrete helicity $\lambda$, positive energy and massless particle, discrete helicity $\lambda$ but negative energy, respectively. Let us take $\qty(\mathcal{H}_1,\rho_1)$ as the representation given by \eqref{eq_massless_irrep} with $\mathcal{H}_1 = L^2 \left(X_0,\frac{d^3p}{|\vec{p}|}\right)$. Let us consider the representation $\qty(\mathcal{H},\rho)$ with  $\mathcal{H} = \mathcal{H}_1 \oplus \mathcal{H}_2$
 and $\rho = \rho_1 \oplus \rho_2$.
We can write $\rho$, in an obvious notation,  as
\begin{equation}\label{eq_direct_sum_rep}
    \widetilde{P}_+^\uparrow \xrightarrow[]{\;\;\rho\;\;}  \mqty( \rho_1 \qty(\widetilde{P}_+^\uparrow) & 0 \\ 0 & \rho_2 \qty(\widetilde{P}_+^\uparrow)).
\end{equation} 
From \eqref{eq_direct_sum_rep}, we can construct the Lie algebra representation of $\widetilde{P}_+^\uparrow$. This is trivial, and we only write explicitly the generator of time translation (the energy), using the unitary transformation $V$ given by \eqref{eq_V_transformation}, in the representation $L_p^2(\mathbb{R}^3)\oplus L_p^2(\mathbb{R}^3)$:
\begin{equation}
    \mathbb{H} = \mqty( p^0 & 0 \\ 0 & -p^0 ),
\end{equation}
where $p^0 = \sqrt{\vec{p} ^2}$. 
 
Now, we analyze the Weyl equation introducing the operator $\mathbb{H}_{w}$ (Weyl Hamiltonian):
\begin{equation}\label{eq_Weyl_Hamiltonian}
\mathbb{H}_{w} \Psi (\vec{x})= -i \qty(\sigma_1 \partial_{x_1} + \sigma_2 \partial_{x_2} + \sigma_3 \partial_{x_3})  \Psi (\vec{x}),
\end{equation}
for all $ \Psi \in \mathcal{D}(\mathbb{H}_{w}) \subset L^2(\mathbb{R}^3)\oplus L^2(\mathbb{R}^3)$, where $\{\sigma_i\}_{i=1}^3$ are linearly independent $2\times 2$ matrices which satisfy the anticommutation relation  $\sigma_i\sigma_j+\sigma_j\sigma_i=2\delta_{ij} \mathbbm{1}_2$ for $i,j=1,2,3$ and  $\mathcal{D}(\mathbb{H}_{w})$ being the standard domain (similar to the Dirac equation \cite{Thaller1992}) where $\mathbb{H}_{w}$ is selfadjoint. 

Using de Pauli matrices, the equation \eqref{eq_Weyl_Hamiltonian} is written as:
\begin{equation}
 \qty(\mathbb{H}_{w} \Psi) (\vec{x}) =
\mqty(
        -i  \partial_{x_3} & -i  \partial_{x_1} - \partial_{x_2} \\ -i  \partial_{x_1} + \partial_{x_2} &  i  \partial_{x_3}
       )\Psi (\vec{x}).
\end{equation}

The evolution equation with this Hamiltonian (Weyl equation) is given by:
\begin{equation}\label{eq_Weyl_evolution}
 i  \frac{\partial \Psi (\vec{x},t) }{\partial t} = \mathbb{H}_{w} \Psi (\vec{x},t),
\end{equation}
where,
\begin{equation}
\Psi (\vec{x}) = \mqty( \psi_1(\vec{x}) \\
    \psi_2(\vec{x}) ) \in  \mathcal{D}\qty(\mathbb{H}_w).
\end{equation}

Now, we will see that \eqref{eq_Weyl_evolution} is compatible with a unitary  representation of $\widetilde{P}^\uparrow_+$ which is physically equivalent to the reducible representation $(\mathcal{H},\rho)$.

For obtaining the unitary equivalence of these representations, that we have pointed out, we use the Fourier transformation:
\begin{equation}
 \qty(F\psi_k)(\vec{p}) = \frac{1}{(2\pi)^{3/2}} \displaystyle \int  \psi_k(\vec{x}) e^{-i \vec{x} \cdot \vec{p}} d^3 x,
\end{equation}
for $ k=1,2$ and $\psi_k \in L^2(\mathbb{R}^3)$, which applied to $\mathbb{H}_{w}$ is:
\begin{equation}
F \mathbb{H}_w  F^{-1}=\widetilde{\mathbb{H}}_w =  \mqty(
        p^3 & p^1 - i p^2 \\ p^1 + i p^2 & -p^3).
 \end{equation}
For each $\vec{p}$,  $\widetilde{\mathbb{H}}_w$ is  a hermitian matrix $2 \times 2$, which is not diagonal when $p^1$ or $p^2$ are different from zero, and has the eigenvalues:
\begin{equation}
\begin{array}{rcl}
 E_1(\vec{p}) & = & - E_2 (\vec{p})  \\ & = & \sqrt{(p^1)^2 + (p^2)^2 + (p^3)^2 } = p^0 .
 \end{array}
\end{equation}
We construct the unitary operator $u(\vec{p})$ (with $p^1$ or $p^2$  different from zero) which transforms $\widetilde{\mathbb{H}}_w(\vec{p})$ to its diagonal form:
\begin{equation}
 u(\vec{p}) =  \dfrac{\qty(p^0+p^3)\mathbbm{1}_2 + \sigma_3 \qty(\sigma_1 p^1 + \sigma_2 p^2)}{\sqrt{2p^0\qty(p^0 + p^3)}}.
\end{equation}
Therefore, 
\begin{equation}
 u(\vec{p}) \widetilde{\mathbb{H}}_w(\vec{p}) u^{-1}(\vec{p}) = p^0 \sigma_3  = \mathbb{H},
\end{equation}
with 
\begin{equation}
 u^{-1}(\vec{p}) =  \dfrac{\qty(p^0+p^3)\mathbbm{1}_2 - \sigma_3 \qty(\sigma_1 p^1 + \sigma_2 p^2)}{\sqrt{2p^0\qty(p^0 + p^3)}},
\end{equation}
being the inverse transformation of $u(\vec{p})$.

Thereby, the unitary transformation, from $\mathbb{H}_w$ to its diagonal form in momentum, is composing $u$ with $F$. The operator $W \equiv uF$ converts $\mathbb{H}_{w}$ in a multiplicative operator by the diagonal matrix (in momentum):
\begin{equation}
\qty(W \mathbb{H}_{w} W ^{-1} \widetilde{\Psi})(\vec{p}) = \qty( \mathbb{H} \widetilde{\Psi})(\vec{p}),  
\end{equation}
for all $ \widetilde{\Psi}(\vec{p}) \in L^2_p(\mathbb{R}^3)\oplus L^2_p(\mathbb{R}^3)$. 

Therefore, since $\mathbb{H}$ is the generator of time translation of $\rho_1 \oplus \rho_2$, we have that the Weyl equation is compatible with a representation that is unitarily equivalent, through $W$,  to the direct sum of two massless unitary irreducible representations, of \underline {arbitrary helicity} $ \lambda $ ,  given by (\ref{eq_massless_irrep}) and its corresponding representation with negative energy. Then, the Weyl equation for massless particles (as the Dirac equation for the electron) does not describe an elementary particle. We have to remark that the Weyl equation is compatible with any helicity, but this compatibility is at the level of particles and not in QFT. This result is also true, as we said before, for Maxwell type equation. That is, it is not difficult to prove, following the same lines as with the Weyl equation, that Maxwell type equation is compatible with a representation that is unitarily equivalent  to the direct sum of two massless unitary irreducible representations, of arbitrary helicity $ \lambda $, given by (\ref{eq_massless_irrep})  and its corresponding representation with the same helicity but negative energy.

Finally, we define, the analogue to the Foldy-Wouthuysen transformation of the Dirac electron equation, for the Weyl massless case:
\begin{equation}\label{eq_FW_transformation}
 U_{FW} = F^{-1} W,
\end{equation}
which transforms $\mathbb{H}_{w}$ like:
\begin{align}
\begin{split}
 \qty( U_{FW} \mathbb{H}_{w} U_{FW}^{-1} \Psi)(\vec{x}) &= \left( \begin{array}{cc} \sqrt{ -\Delta} & 0 \\ 0 & - \sqrt{-\Delta} \end{array} \right) \Psi(\vec{x}) \\
 &= \sqrt{-\Delta}\sigma_3 \Psi(\vec{x}).
\end{split}
\end{align}
for all $ \Psi(\vec{x})\in \mathbb{D} \subset L^2_x(\mathbb{R}^3)\oplus L^2_x(\mathbb{R}^3)$. This defines the position representation (obviously, with the restriction to positive energy). Thus, we see that the multiplicative operator by the components of $\vec{x}$ is not the position operator in $L^2(\mathbb{R}^3)\oplus L^2(\mathbb{R}^3)$, where the Weyl Hamiltonian is written by
\eqref{eq_Weyl_Hamiltonian}. The position operator is the multiplicative one in the representation where $\mathbb{H}_{w}$  is given by $\sqrt{-\Delta}\sigma_3$. Therefore, the position operator, in the original representation, given by \eqref{eq_Weyl_Hamiltonian}, comes from the inverse of the Foldy-Wouthuysen transformation [the inverse of \eqref{eq_FW_transformation}]. This is also true for the Maxwell type equation and explains why the multiplicative operator is not the position operator in the Maxwell standard representation (the multiplication by $\vec{x}$ can be not applied to the Maxwell type equation because it destroys the divergence condition). However, in the Foldy-Wouthuysen representation, for the Maxwell type equation, is the multiplicative one.

\section{Discussion}

In this article, we have clarified in what sense massless particles of non-zero helicity are localized. This result applies, in particular, to the two kind of photons (the one with helicity $+1$ and the other one with helicity $-1$) and to the two kind of gravitons (the one with helicity $+2$ and the other with helicity $-2$). To obtain localization, we essentially  demand canonical commutation relations and selfadjointness of position and momentum operators (in a precise way, we ask the fulfillment of the axioms I-V).

Of course, our considerations of localization apply to any elementary particle, because if a particle is localizable in the sense of NWW, it is trivially localizable in the sense of axioms I-V. We think that it is better to have a localization than none (for the case of massless particles of non-zero helicity).  We notice that the position operator \eqref{eq_position_operator}, for any helicity, is the same that the one given by NWW axioms for the massless particle of zero helicity (naturally, the representations in the Hilbert space are different). We emphasize that these results, for instance \eqref{eq_hamiltonian_evolution} and \eqref{eq_position_evolution}, are completely relativistic, in the sense of quantum mechanics, because we are working with unitary irreducible projective representations of the restrictive Poincar\'e group,  according to the Wigner's theorem.

The wave equation of massless particles that we have derived are unique, up to unitary equivalence, from minimal requirements such as  canonical commutation relations, in the Weyl sense, between  position and momentum operators, and the irreducibility of projective representation of the Poincar\'e group corresponding to massless particles. In the literature, the hypothesis to obtain wave equations are not clear, or the equations are postulated. Moreover, canonical commutation relations are sometimes not satisfied, see, e.g.,   \cite{Birula2012}.

Evidently, the localization obtained from axioms I-V could have additional properties that depend on the considered relativistic representation. This is the case, for example, of massive particles that have ``nice" transformations under the rotation subgroup. But, this is not the case, for example, of massless particles of non-zero helicity. In the same way, it could be desirable to have a localization compatible with the classical notions of relativistic localization. However, none representation has these properties \cite{Hegerfeldt1985}. Nevertheless,  it is completely relativistic from the quantum mechanics point of view, and it does not disagree with the well known non-local nature of quantum mechanics.

The importance of position representation could be underrated by the abstract Hilbert space approach (since it is another representation more). However, it is very important because approximations, from the physical point of view, can be easier and more visible to formulate in position representation than in other representations. Approximations appear everywhere in physics. Moreover, it allows the use of functional analytic methods that could not be clear on the abstract Hilbert approach. 

As it is well known, QFT has associated particles with it. Therefore, the massless particles that appear in this work could be considered as part of a QFT. Let us notice, however, that the symmetries of QFT are not necessarily the symmetries of the irreducible representations that describe its particles. For instance, if a QFT is invariant under parity, this parity operator acts on the Fock space of the theory and does not act necessarily on the Hilbert space of one particle. The so called ``first quantization space", from which the Fock space is constructed, is not necessarily the space of one particle but, it could be, a direct sum of two different particles. Then, if we are considering the physics of one particle, it is an unnecessary connecting it with QFT. This situation is clear  after showing that Weyl equation is compatible with any helicity, like the Maxwell type equation, or like the Dirac equation, for the case of one particle. Nevertheless, this compatibility with any helicity will not be necessarily true if we take into account QFT. The essence of this work has been the problem of the quantum mechanical description of one massless particle of arbitrary helicity, then the creation and destruction of particles (QFT) does not appear in our description. Of course, we cannot deny the connection between particles and QFT.

\begin{acknowledgments}
The present work is an extended version of some parts of S. Tabban's master thesis at Universidad Central de Venezuela, that was made  under the supervision of Prof. Abraham Lozada, in 2015.

\end{acknowledgments}

\end{document}